\documentclass[twocolumn,aps,prl,preprintnumbers,showpacs,nofootinbib]{revtex4}
\usepackage[utf8]{inputenc}
\usepackage[english]{babel}
\usepackage{amsmath}
\usepackage{amsfonts}
\usepackage{amssymb}
\usepackage{epsfig}
\usepackage{graphics,psfrag,rotating}
\usepackage{graphicx}
\usepackage{dcolumn}
\usepackage{bm}
\usepackage{epstopdf}
\usepackage{color}
\usepackage[usenames,dvipsnames,svgnames]{xcolor}
\usepackage[colorlinks=true,
            linkcolor=red,
            urlcolor=gray,
            citecolor=blue]{hyperref}
  \usepackage{hyperref}

\newcommand{\lsim}   {\mathrel{\mathop{\kern 0pt \rlap
{\raise.2ex\hbox{$<$}}}
 \lower.9ex\hbox{\kern-.190em $\sim$}}}
\newcommand{\gsim}   {\mathrel{\mathop{\kern 0pt \rlap
{\raise.2ex\hbox{$>$}}}
\lower.9ex\hbox{\kern-.190em $\sim$}}}

\def\3nab{\tilde{\nabla}}

\def\hsp5{\hspace{5mm}}

\def\case#1/#2{\textstyle\frac{#1}{#2}}

\def\ber {\begin{eqnarray}}
\def\eer {\end{eqnarray}}
\def\bea {\begin{eqnarray}}
\def\eea {\end{eqnarray}}

\def\ne {n_{\!_E}}

\def\bc {\begin{center}}
\def\ec {\end{center}}
\def\case#1/#2{\frac{#1}{#2}}

\newcommand{\bw}{\begin{widetext}}
\newcommand{\ew}{\end{widetext}}

\newcommand{\be}{\begin{equation}}
\newcommand{\bse}{\begin{subequation}}
\newcommand{\ese}{\end{subequation}}
\newcommand{\ee}{\end{equation}}
\newcommand{\eei}{\end{eqnarray}\indent\indent}
\newcommand{\ba}{\begin{array}}
\newcommand{\ea}{\end{array}}
\newcommand{\bal}{\begin{eqnarray}}
\newcommand{\eal}{\end{eqnarray}}

\def\case#1/#2{\textstyle\frac{#1}{#2} }

\newcommand{\bver}{\begin{verbatim}}
\newcommand{\ever}{\end{verbatim}}

\def\prd{{\em Phys. Rev.\/} {\bf D}}

\begin{document}


\title{The Raychaudhuri equation in homogeneous cosmologies}


\author{F.\,D. Albareti$^{(a,b)\,\footnote{E-mail: fdalbareti@ucm.es}}$}
\author{J.\,A.\,R. Cembranos$^{(b)\,\footnote{E-mail: cembra@fis.ucm.es}}$}
\author{A. de la Cruz-Dombriz$^{(b)\,\footnote{E-mail: dombriz@fis.ucm.es} 
}$}
\author{A. Dobado$^{(b)\,\footnote{E-mail: dobado@fis.ucm.es}}$}
%
\affiliation{$^{(a)}$ Departamento de F\'isica Te\'orica, Universidad Aut\'onoma de Madrid, Campus de Cantoblanco, E-28049 Madrid, Spain}
\affiliation{$^{(b)}$Departamento de F\'{\i}sica
Te\'orica I, Universidad Complutense de Madrid, Ciudad Universitaria, E-28040 Madrid,
Spain}
%
%

\date{\today}


%
%

\begin{abstract}

In this work
we address the issue of studying the conditions required to
guarantee the Focusing Theorem for both null and timelike geodesic congruences
by using  the Raychaudhuri equation. In particular we study the case of Friedmann-Robertson-Walker as well as more general Bianchi Type I spacetimes.
The fulfillment of the Focusing Theorem is mandatory in small scales since it accounts for the attractive character of gravity. However, the Focusing Theorem
 is not satisfied at cosmological scales due to the measured negative deceleration parameter.  The study of the conditions needed
for congruences convergence is not only relevant at the fundamental level but also to derive the
viability conditions to be imposed on extended theories of gravity describing the different expansion regimes of the universe. We illustrate this idea for $f(R)$ 
gravity theories.

\end{abstract}

\maketitle



\section{Introduction}
\label{intro}


The accelerated expansion of the Universe may be studied using congruences of geodesics which provide a coordinate-independent description of the phenomenon.
A useful tool to analyze the behavior of geodesics is the Focusing Theorem (FT). This theorem is based on the so called Raychaudhuri  equation (also independently found by Landau \cite{Sayan}). Assuming the Einstein's equations and using energy conditions it is established that timelike and null geodesics if initially converging will focus until zero size in a finite time. Nevertheless, the requirements for the FT to hold are not satisfied in the present Universe since the deceleration parameter is negative \cite{SIa}.

However, there still exists focusing for certain timelike congruences \cite{ACD} even if the sufficient conditions for the FT are not fulfilled. In any case, one needs to take into account that the conditions on the energy-momentum tensor depend on the particular gravitational theory under study. Indeed, there has been in the last years a large number of proposed gravitational theories \cite{Mod_Grav_Theories}, such as Lovelock models \cite{Lovelock}, Gauss-Bonnet theories \cite{GB}, extra dimensional geometrical models \cite{XD}, disformal gravity \cite{disformal}, Lorentz violating and CPT breaking models \cite{LV}, supergravity \cite{sugra}, non-local gravitational theories \cite{NL}, scalar-tensor models \cite{ST}, or higher spin-tensor models \cite{VT}. We shall not study these particular extensions of the gravitational interaction. On the opposite, we shall show a general study which shall be not constrained to Einstein gravity. We shall also pay particular attention to the so-called $f(R)$ theories of gravity \cite{fR} as a simple example beyond the standard gravitational theory.

The aim of this paper is twofold: On one hand, we shall derive the sufficient conditions to the FT holds in the context of $f(R)$ theories assuming the usual energy conditions imposed on the energy-momentum tensor. The conditions derived must be satisfied at small scales for the models under consideration. On the other hand, using these conditions one may restrict the parameter-space for classes of $f(R)$ models taking into account the expected
spacetime contribution to the Raychaudhuri equation at a cosmological scales and the present acceleration of the Universe.

In order to illustrate the relevance of our analysis, we shall apply our results to flat Friedmann-Robertson-Walker (FRW) models and to homogeneous anisotropic models
Bianchi Type I. In this realm we shall consider both GR as well as the most common modified gravity theories, the so-called $f(R)$ theories which have drawn a lot of attention in the last years.
For these theories the gravitational action in the metric formalism  can be written as
\begin{equation}
\label{lagr f(R)}
\mathcal{A}=\frac{1}{16\pi G }\int {\rm d}^4 x \sqrt{-g}\left[R+f(R)\right]\;,
\end{equation}
where $R$ is the Ricci scalar and  $f$ is a general differentiable
(at least $C^2$) function of the Ricci scalar.
In the metric formalism, the modified Einstein equations (EFEs) derived by varying this action with respect to the metric are usually fourth order.


Throughout this communication, our conventions are
\begin{eqnarray}
g_{ab} \sim \left(-,+,+,+\right)\,,
\end{eqnarray}
\begin{eqnarray}
R_{abc}^{\;\;\;\;\;\;d}\equiv \partial_b \Gamma^d_{ac}-\partial_a \Gamma^d_{bc}+\Gamma^e_{ac}\Gamma^d_{eb}-\Gamma^e_{bc}\Gamma^d_{ea}\,,
\end{eqnarray}
\begin{eqnarray}
R_{ac} \equiv R_{abc}^{\;\;\;\;\;\;b}\,.
\end{eqnarray}
Finally, in the following we shall adopt geometrized units $c=G=1$.


\section{The Raychaudhuri equation and the Focusing Theorem}


Given a congruence of timelike geodesics, its tangent vector field ${\boldsymbol \xi}$
and the proper time $\tau$ of an observer in free fall along one of these geodesics, the Raychaudhuri equation can be expressed as \cite{Wald,Raychaudhuri}
\begin{eqnarray}
\frac{\text{d}\theta}{\text{d}\tau}=-\frac{1}{3}\,\theta^2-\sigma_{ab}\sigma^{ab}+\omega_{ab}\omega^{ab}-R_{ab}\xi^a\xi^b\,,
\label{ray}
\end{eqnarray}
where $\theta$, $\boldsymbol{\sigma}$ and $\boldsymbol{\omega}$ are respectively the expansion, shear and twist which are the trace, trace-less symmetric part and antisymmetric part respectively of the covariant derivative of the vector field $\boldsymbol{\nabla} \boldsymbol{\xi}$. This equation may be extended to null geodesics \cite{Wald,Sachs}. 
The Raychaudhuri equation can be thought of as geometrical identities which do not depend on any gravitational theory. They provide the evolution of the expansion of a geodesic congruence.

The term $-R_{ab}\xi^a\xi^b$ encapsulates the contribution of spacetime geometry and it does not depend on the derivatives of the vector field. Thus this terms possesses more general implications than the other terms in (\ref{ray}). It has a geometrical interpretation as a mean curvature
\footnote{This mean curvature does not correspond to the one defined in classical differential geometry of surfaces, indeed it has no relation with a surrounding space.}
\cite{Eisenhart,ACD} in the direction of $\boldsymbol{\xi}$. Since this interpretation is independent of the conventions used to define the geometrical tensors, we shall deal with this mean curvature in the present discussion
\begin{eqnarray}
{\cal M}_{\boldsymbol{\xi}} &\equiv& -R_{ab} \xi^a \xi^b\,, \\
{\cal M}_{\boldsymbol{k}} &\equiv& -R_{ab} k^a k^b;
\end{eqnarray}
where we have extended the notation to null vectors even though this case lacks analogous interpretation
as a mean curvature.

If one is interested in the focusing of geodesics, congruences without vorticity $\omega_{ab}=0$ must be considered in order to avoid the presence of centrifugal forces. It follows that the only term that may be positive in the Raychudhuri equation (\ref{ray}) is ${\cal M}_{\boldsymbol{\xi}}$. Information about the sign of this contribution may be obtained from observation of the deceleration parameter \cite{ACD}. In this reference authors found
that the mean curvature contribution is positive for timelike vectors lying inside a certain cone around the cosmic time vector of Friedmann-Robertson-Walker models. Therefore, the FT is not satisfied in the present Universe at a cosmological scale.
From a dynamical point of view it is possible to use gravitational field equations and assume the usual energy conditions on the energy-momentum tensor to obtain information about ${\cal M}_{\boldsymbol{\xi}}$. This approach was carried out in the context of $f(R)$ theories in \cite{ACDD}.

The usual energy conditions considered are
%
\begin{eqnarray}
T_{ab}\xi^a\xi^b &\geq 0 \hspace{1.5cm} \text{WEC,}\,\\
\label{WEC}
T_{ab}\xi^a\, T^{bc}\xi_c &\leq 0 \hspace{1.5cm} \text{DEC,}\,\\
\label{DEC}
T_{ab}\xi^a\xi^b + \frac{1}{2}T &\geq 0 \hspace{1.5cm} \text{SEC,}\,\\
\label{SEC}
T_{ab} k^a k^b &\geq 0 \hspace{1.5cm} \text{NEC}\,.
\label{NEC}
\end{eqnarray}

%
The first one dubbed weak energy condition (WEC) states that the energy density measured by any observer must be positive. 
The dominant energy condition (DEC) ensures a future directed timelike flux of energy-momentum ($-T_{ab}\xi^a$). 
Recently, the requirement for a timelike flux of energy-momentum has been referred to as the flux energy condition \cite{Prado}.
The third one, strong energy condition (SEC) is also accomplished by almost all classical system \cite{HE} and it can be seen that in General Realtivity (GR) it implies ${\cal M}_{\boldsymbol{\xi}} \leq 0$ for every timelike vector. The last one, null energy condition (NEC), must be satisfied by continuity if
either the WEC or SEC are assumed. The SEC and NEC are the necessary conditions in GR for the Focusing Theorem to hold. Since almost all classical system
satisfy the SEC there is no reason to think that  the FT is not satisfied in GR. However, the more general field equations of $f(R)$ theories allow the violation of FT
even if the SEC is assumed \cite{ACDD}.


In the following we will study  geodesic congruences with  vanishing  vorticity ($\omega_{ab}= 0$) and we will consider both
time-like congruences $\xi^a =(\xi^0,\xi^i)$
with $\xi_a\xi^a=-1$
and null congruences $k^a=(k^0,k^i)$
with $k_a k^a=0$. For both types of geodesic congruences, if the Convergence Condition (CC)
 ${\cal M}_{\boldsymbol{\xi}}  \le 0$ or
${\cal M}_{\boldsymbol{k}} \le 0$ is satisfied, the Raychaudhuri equation \cite{Wald}  implies respectively
\begin{eqnarray}
 \frac{{\rm d} \theta}{{\rm d} \tau} &\le& -\frac{1}{3} \theta^2\,,\\
\frac{{\rm d} \tilde \theta}{{\rm d} \lambda} &\le& -\frac{1}{2} \tilde \theta^2\,,
\end{eqnarray}
where $\tau$ is the proper time for the timelike geodesic congruence,
$\lambda$ is an affine parameter for the null geodesic congruence and $\theta$ and  $\tilde \theta $
hold for the expansions corresponding to each congruence respectively.
As a straightforward consequence of both inequalities we conclude that, provided the congruence is initially converging ($\tilde \theta_0 <0$, $\theta_0 <0$),
it will shrink to zero size in a finite time. Conversely, if the congruence is expanding, it was focused to zero size in a finite time in the past.  In the following we shall refer to this result as the Focusing Theorem (FT).


\section{Friedmann-Robertson-Walker spacetimes}
\label{Sec:RW}

In this section we shall study
the timelike and null geodesics evolution in spatially flat ($k=0$) Friedmann-Robertson-Walker (FRW) models  whose relevance in cosmological scenarios is manifest.
For these models, the line element in cosmic time can be written as
\be
 {\rm d}s^2\,=\,-{\rm d}t^2+a^2(t)\,{\rm d}\vec{r}^{\,2}
\label{metric_FRW}
\ee
where as usual $a(t)$ holds for the cosmological scale factor and $t$ for the cosmic time.
%
%

\subsection{Timelike Geodesics}


For  the metric in \eqref{metric_FRW}, timelike geodesics parametrized by the proper time have a unitary tangent vector which can always be
written as
\be
\xi^a=\gamma (1,v^i)	\,,
\label{tvector}
\ee
with
\be
\gamma\equiv \frac{1} {\sqrt{1-a^2 v^2 }}\,,
\label{gamma_def}
\ee
and $v^2=\sum_i (v^i)^2$, ($i=1,2,3$).
The geodesic equation is just
 \be
 \frac{d}{d t}(\gamma v^i) + \gamma H v^i=0
 \ee
 where $H=\dot{a}/a$ is the Hubble parameter. The solution for any $t_0  \le t$ is given in terms of the corresponding quantities evaluated at $t=t_0$  as
 \be
  \gamma  v^i a^2=\gamma_0  v^i_0  a_0^2
  \label{integral_motion}
 \ee
so that the tangent vector of the geodesic timelike congruence can always be
 written as
\be \xi^{a}\,=\,\left(\sqrt{1+\frac{\gamma_0^2  v_0^2  a_0^4}{a^2}},\gamma_0
v^i_0 \frac{a_0^2}{a^2}\right).
\ee
Notice that we are considering, here and in the following, a spatial region small enough so that the velocity field can be considered as constant at some given time. In particular $v^i_0$ is locally point independent so we can neglect its spacial gradient.
 Now the CC is
 \be
 \frac{1}{1-a^2v^2} \left[(3-a^2v^2) \frac { \ddot{a}   }{a} -2 v^2 a^2\frac{\dot{a}^2 }{a^2}    \right] \le 0
 \ee
which for geodesic congruences can be written as
 \be
     a \ddot{a}  \le\frac{1}{1+\frac{3a^2}{2\gamma_0^2 v_0^2 a_0^4}} \  \dot{a}^2\,.
     \label{CCtime_2_JAR}
 \ee
%
%
The previous expression can be rewritten in terms of redshift $z=a^{-1}-1$,  the 
deceleration parameter 
$q\equiv - \frac{1}{H^2} \frac{\ddot{a}}{a}$ and the first integral of  motion (\ref{integral_motion}) as
\begin{eqnarray}
&&q(z)+\left\{1+\frac{3}{2}\left[\frac{1+z_0}{v_0(1+z)}\right]^2\left[(1+z_0)^2-v_0^{2}\right]  \right\}^{-1}\, \geq\, 0\nonumber\\
&&
\label{CCtime_2}
\end{eqnarray}
where $z_0$ holds for the redshift at which the integral of motion (\ref{integral_motion})  is evaluated.
%
%


\subsection{Null Geodesics}


For null geodesics the CC just renders
\be
(k^0)^{2}\left(\frac{ \ddot{a}   }{a} -  \frac{\dot{a}^2 }{a^2}    \right) \leq 0\,.
 \label{null}
\ee
Therefore the CC condition \mbox{${\cal M}_{\boldsymbol{k}}<0$} leads to the simple  inequality
\be
   a \ddot{a}  \le \dot{a}^2,
\ee
or alternatively, using the definition of the deceleration parameter
the CC condition is just
\be
 q  \ge -1\,,
 \label{CCnull}
\ee
which guarantees the FT for null geodesics.  The limiting case saturating the previous inequality is given by the exponential expansion (flat de Sitter Universe)
\be
a(t)=a_0\, {\rm e}^{H_0(t-t_0)}.
\ee
Thus we may conclude  that this CC condition is less restrictive and
more easily satisfied by null geodesics than by timelike geodesics.



\section{Anisotropic models}
\label{Section_6}

In this section we extend the previous discussion by considering anisotropic spaces, in
particular Bianchi Type I  spaces which can always be described by the
metric
\be
{\rm d}s^2 = -{\rm d}t^2+\sum_i a_i^2({\rm d}x^i)^{2}\,.
\ee
For this kind of geometry the  Ricci tensor is diagonal and the non-vanishing components are
\be
R_{00}=-\sum_i  \frac{\ddot a_i}{a_i}\;\;,\;\;
 R_{ii}=a_i^2\left[ H_i \left(3H - H_i\right)+\frac{\ddot a_i}{a_i}\right]
 \ee
 where
 \be
 H_i\equiv\frac{\dot{a}_i}{a_i}\;\;,\;\;
H\equiv\frac{1}{3}\sum_i H_i
 \ee
are the partial Hubble parameters and the average Hubble parameter respectively. Obviously FRW spacetime can be straightforwardly
obtained from these more general metrics as the particular case $a_i(t)= a(t)$ for all spatial indices.

Einstein equations (\ref{1}) can be solved for these metrics for different kind of fluids. For example for dust matter, i.e.,  $p=0$,
it is possible to find the solution
\be
a_i(t)=a(t)^{(1-2\lambda_i)}\left(\frac{t}{t_0}\right)^{\frac{4\lambda_i}{3}}
\ee
where
\be
\lambda_i\equiv\frac{1}{2}\left(1-\frac{\log a_{0i}}{\log a_0}\right)
\ee
and
\be
a(t)^3 \equiv \prod_i a_i(t)=\alpha \frac{t}{t_0}\left( \frac{t}{t_0}+\beta\right)
\ee
which implies $\lambda \equiv \sum_i \lambda_i=0$. The Einstein equations (\ref{1}) are fulfilled provided that
\be
1+\beta = \frac{1}{6 \pi \rho_0 t_0^2}.
\ee
On the one hand, for $\beta$ different from zero, the condition $\lambda'^2=\sum_i \lambda^2_i= 3/2$ needs to be satisfied. On the other hand, for $\beta = 0$ there exists a solution with $\lambda_i=0$ (for all $i$)
which corresponds to the FRW spacetime with
\be
a_i(t)=a(t)=a_0\left(\frac{t}{t_0}\right)^{2/3}
\ee
which corresponds to the well known Einstein-de Sitter universe.

\subsection{Timelike geodesic congruences}

 For the general Bianchi Type I spaces the tangent vector of the geodesic congruenses $\xi^a$ can be expressed again as $\xi^a=\gamma(1, v^i)$
where
\be \gamma = \frac{1}{\sqrt{1-\sum_i v_i^2 a_i^2}}
\label{eq_1_timelike}, 
\ee
and $v_i \equiv   v^i$.  
Thus the geodesic timelike equation reads
\be
 \frac{{\rm d}}{{\rm d} t}(\gamma v_i) + \gamma  H_i v_i=0.
 \label{eq_2_timelike}
\ee
This is a coupled system of differential equations for the velocity components
$v_i$ whose solution is
\be v_i^2
 = \frac{c_i^2}{\gamma^2a_i^4}
 \ee
  where
  \be
\gamma^2=1+\sum_i \frac{c_i^2}{a_i^2}
\ee
 and the $c_i$ are given  by the
initial conditions as $c_i\equiv \gamma_0 v_{0i}a_{0i}^2$. Therefore the tangent vector to the geodesic timelike congruence in these spaces can be written as
\be
\xi^a=\left(\sqrt{1+ \sum\frac{\gamma_0^2v_{0i}^2a_{0i}^4}{a_i^2}},  \gamma_{0i} v_{0i}\frac{a_{0i}^2}{a_i^2}\right)
\ee
and then it is very easy to check $\xi^2=\xi^a \xi_a=-1$. For these  timelike congruences the CC reads
\begin{eqnarray}
 {\cal M}_{\boldsymbol{\xi}} &=&\frac{1}{1-\sum_j a_j^2v_j^2} \nonumber\\
  &&\times\sum_i \left[(1-a_i^2v_i^2) \frac { \ddot a_i   }{a_i} + v_i^2 a_i^2H_i(H_i-3H   )\right] \le 0\nonumber\\
  &&
\end{eqnarray}
which after substituting the aforementioned partial results for geodesic curves becomes
\begin{eqnarray}
 {\cal M}_{\boldsymbol{\xi}} &=& \sum_i \Big[\left(1+\sum_j \frac{\gamma_{0}^2v_{0j}^2a_{0j}^4}{a_j^2}-\frac{\gamma_{0}^2v_{0i}^2a_{0i}^4}{a_i^2}\right) \frac { \ddot a_i   }{a_i}\nonumber\\
  &+& \frac{\gamma_{0}^2v_{0ij}^2a_{0i}^4}{a_i^2}H_i(H_i-3H)\Big] \le 0\,.
\end{eqnarray}

\subsection{Null geodesic congruences}

For the case of  null geodesic  congruences  the tangent vector can be written as $k^a=(k^0,
k^i)$, with $k^i \equiv k_i$ satisfying  the equation \be k^a D_a k^b
=0 \ee where $k^a= {\rm d}x^a/{\rm d} \lambda$ and
$x^a(\lambda)=(x^0(\lambda),x^i(\lambda))$ are the coordinates of the  points on the congruence
parametrized by the affine parameter $\lambda$.
The general solution for this geodesic equation is given by
\be
 k^a=\left(\sqrt{\sum_j k_{0j}^2a_{0j}^2}\,,\, k_{0i}\frac{a_{0i}}{a_i}\right)
 \ee
so that $k^a k_a=0$.

Now the CC for null congruences reads
\begin{eqnarray}
 {\cal M}_{\boldsymbol{k}} &=&(k^0)^2\left\{\sum_i \frac{ \ddot a_i   }{a_i} -\sum_i  k_i^2 a_i^2[H_i(3H-H_i) ] +\frac{\ddot a_i}{a_i}   \right\}  \nonumber\\
 &&\le 0
\end{eqnarray}
which, in the particular case $a_i(t)=a(t)$, leads us to the standard CC expression for null geodesic in FRW spaces.
For geodesic congruences the previous equation can be rearranged as
\begin{eqnarray}
&& {\cal M}_{\boldsymbol{k}} =\left(\sum_j k_{0j}^2a_{0j}^2\right)\nonumber\\
  &&\times\left\{\sum_i \frac{ \ddot a_i   }{a_i} -\sum_i  k_{0i}^2a_{0i}^2[H_i(3H-H_i) ] +\frac{\ddot a_i}{a_i}   \right\} \le 0\nonumber\\
  &&
\end{eqnarray}
which is a function only on the initial conditions $k_{0i }$ and the scale factors $a_i(t)$ and their first two time derivatives. 

\section{General case in Power-law spacetimes}

In order to see how it is possible to apply the above CC leading to the FC in some particular cases  it is interesting to study some soluble models that can be managed analytically, still not being trivial at all. In the following we will concentrate in the so called power law spaces where the $a_i(t)$ are just
\be
a_i(t)=a_0\left(\frac{t}{t_0}\right)^{\alpha_i}.
\ee
Then the  Ricci tensor components are given by
 \be
R_{00}=-\frac{1}{t^2}(\alpha-\alpha'^2)\,,
 \ee
 and
 \be
 R_{ii}=-\frac{a_{0i}^2}{t^2}\left(\frac{t}{t_0}\right)^{2\alpha_i}
 \alpha_i(\alpha-1)\,,
 \ee
  where $\alpha\equiv \sum_i
\alpha_i$ and $\alpha'^2 \equiv \sum_i \alpha_i^2$. Therefore the scalar
curvature is simply given by
 \be
 R=\frac{1}{t^2}
(\alpha'^2+\alpha^2-2\alpha)\,,
\label{Ricci_power_law}
 \ee
 and the element of spatial
volume is
\be {\rm d}V=\prod_i a_i \left(\frac{t}{t_0}\right)^{\sum\alpha_i}{\rm d}x^i\,=\, \left(\frac{t}{t_0}\right)^{\alpha}\prod_i a_i {\rm d}x^i. \ee
From the above
expression (\ref{Ricci_power_law}) it is obvious that the conditions $\alpha=1$ and
$\alpha'^2=1$ are equivalent to $R_{ab}=0$, i.e., Kasner spaces. Thus
Kasner spaces are vacuum solutions for GR (without cosmological
constant) as well as for any $f(R)$ theory with $f(0)=0$. Nontheless, let us remark at this stage that those solutions may
evolve in time non trivially.
For $\alpha$ and $\alpha'^2$ different from $1$, these
homogeneous power-law spaces may represent vacuum solutions of GR or
may account for other solutions in the context of more general $f(R)$ gravities.

 In the general case the CC for null geodesics in power-law spaces yields
\begin{eqnarray}
 {\cal M}_{\boldsymbol{k}}&=&\frac{1}{t^2}\Big[\left(\sum_i a_{0i}^2k_{0i}^2\right)(\alpha'^2-\alpha)\nonumber\\
&-&(\alpha-1)\sum_i \alpha_i  a_{0i}^2k_{0i}^2\Big] \le 0\,,
\end{eqnarray}
whereas for timelike geodesics the corresponding CC condition reads
\be
 {\cal M}_{\boldsymbol{\xi}}  = \gamma^2(\alpha'^2-\alpha)  - \sum_i\alpha_i(\alpha-1)\gamma_0^2v_{0i}^2a_{0i}^2\left(\frac{t_0}{t}\right)^{2\alpha_i}   \le 0.
\ee
Obviously the CC is fulfilled in both cases whenever $\alpha=\alpha'^2 =1$ since, as stressed above, this requirement implies $R_{ab}=0$. In order to see how these conditions work, it is of particular importance to consider the simple situation in which the expansion (or contraction) happens only in one spatial direction. Then it is possible to choose the coordinates so that
\be
{\rm d}s^2\,=\,-{\rm d}t^2+a_{01}^2\left(\frac{t}{t_0}\right)^{2\alpha}{\rm d} x_1^2+{\rm d} x_2^2+{\rm d} x_3^2\,,
\ee
i.e., we can choose $\alpha_2 =\alpha_3= 0$ and consequently $\alpha_1=\alpha$ and $\alpha'^2 = \alpha^2$ and $a_{02}=a_{02}=1$.  The CC for null  geodesics becomes in this space
\be
 {\cal M}_{\boldsymbol{k}}\,=\,\frac{1}{t^2}\alpha(\alpha-1)k_{T0}^2  \le 0\,,
\ee
\\
where $k_T^2=k_2^2+k_3^2$. For $k_{T0}^2  \ne 0$ the above condition reduces to $\alpha(\alpha-1) \le 0$ which is equivalent to $\alpha \in [0,1]$. Thus only in this case, or when  $k_{0T}^2=0$ we have CC and FT. Notice that in the cases $\alpha < 0$ or $\alpha > 1$ we have $R >0$. For the timelike
congruences the CC is
\be
 {\cal M}_{\boldsymbol{\xi}} \,=\, \alpha(\alpha-1)(1+\gamma_0^2v_{T0}^2)   \le 0\,.
\ee
Therefore, for any initial conditions, only for $\alpha \in [0,1]$ both CC and FT are accomplished.  Thus these spaces provide paradigmatic examples of power-law spaces where the  CC is not fulfilled whenever $\alpha$ is smaller than $0$ or bigger than $1$, both for null or timelike  geodesics (provided $k_T$ does not vanish for null geodesics).


\section{General Relativity revisited }
\label{Section_2}


As widely known, the Einstein equations can be written as
\be
R_{ab}-\frac{1}{2}  R\, g_{ab} = 8 \pi \, T_{ab}\,,
\label{1}
\ee
where $T_{ab}$ is the matter energy-momentum tensor. These equations in particular imply $R=-8 \pi \,T$, where $R $ and $T$ represent the traces of the Ricci and energy-momentum tensors respectively. Thanks to the Bianchi second identity, the energy-momentum tensor is covariantly conserved, i.e., $\nabla^a T_{ab}=0$ yielding to the matter continuity equations. This fact may also be interpreted from variational principles as a consequence of the invariance of the Hilbert-Einstein action (or the matter action) under diffeomorphisms, i.e. coordinate transformations, which is commonly referred to as gauge invariance.  From \eqref{1} it is clear that provided matter fulfills the NEC
(\ref{NEC}), 
then ${\cal M}_{\boldsymbol{k}} \le 0$ is automatically satisfied. In other words, in GR with no cosmological constant once the NEC is fulfilled, both the CC and the FT are guaranteed for null geodesic congruences.  In the case of timelike congruences the SEC (\ref{SEC}), 
produces the same result, namely ${\cal M}_{\boldsymbol{\xi}} \le 0$.

Equations \eqref{1} can be generalized by adding a non-vanishing cosmological constant $\Lambda$. Thus the field equations become
\be
R_{ab}-\frac{1}{2} R  g_{ab}  + \Lambda g_{ab} = 8 \pi \, T_{ab}\,.
\label{2}
\ee
 In a cosmological context, one usually assumes that matter can be described by a perfect fluid
 with $\rho$ and $p$ being the density and the pressure measured by a comoving observer and $\boldsymbol{\xi}$ being the fluid congruence four velocity.  The cosmological constant may then be understood as some exotic fluid 
with energy density and pressure given by $\rho_\Lambda = - p_\Lambda= \Lambda/(8 \pi)$. This fluid 
fulfills the NEC but not the SEC. Then, one is led to conclude that in absence of ordinary matter, the FT for null geodesics is accomplished but not for
timelike ones for a positive cosmological constant. This result can be generalized to $f(R)$ theories regardless the $f(R)$ model if one considers constant scalar curvature spacetimes \cite{ACDD}.


\section{$f(R)$ gravity theories}
\label{Section_3}

This kind of theories are defined by adding an arbitrary function $f(R)$ to the Einstein-Hilbert gravitational Lagrangian
which is just proportional to $R$. In this scenario, the field equations in the metric formalism become
\be
(1+f')R_{ab}-\frac{1}{2}  g_{ab} ( R+f) - D_{ab}f'= 8 \pi  \ T_{ab}\,,
\label{fieldeq}
\ee
where $f'= {\rm d} f/{\rm d} R$, and $D_{ab} \equiv \nabla_a \nabla_b - g_{ab} \Box$. The corresponding trace equation is given by
\be
R=\frac{2(R+f)-3 \Box f'+ 8 \pi T}{1+f'}\,.
\label{trace}
\ee
Note at this stage that the case of a cosmological constant can be easily accommodated just by taking $f=-2 \Lambda$.

In the following, let us assume the viability condition
\be
1 + f'(R)   > 0\,.
\ee
This condition is imposed in order to ensure a positive effective gravitational constant $G_{eff} \equiv G/\left(1+f'(R)\right)$. It
states that the main part of the contribution to the Einstein's equations conserves the sign \cite{Pogosian}. This condition also guarantees the non-tachyonic character of the standard graviton.


\subsection{Constant curvature analysis}


One  simple but important scenario appears  when the scalar curvature is constant \mbox{$R=R_0$}. Notice that such scenario includes Minkowski, Schwarzschild and \mbox{(anti-)} de Sitter spacetimes. Then the trace equation (\ref{trace}) becomes
\be
R_0=\frac{2(R_{0}+f(R_0))+ 8 \pi T}{1+f'(R_{0})}\,.
\label{trace_R0}
\ee

In this scenario, it is obtained for null geodesics
\be
 {\cal M}_{\boldsymbol{k}}= -\frac{8 \pi }{1 + f'(R_{0})} T_{ab}k^a k^b\,,
\ee
therefore the assumption of the NEC leads straightforwardly to the accomplishment of the FT, including the case of a positive cosmological constant.

On the other hand, for timelike congruences we get
\be
{\cal M}_{\boldsymbol{\xi}} =\frac{1}{1+f'(R_{0})} \left[\frac{1}{2}\left(R_0+f(R_{0})\right)- 8 \pi  T_{ab}\xi^a \xi^b\right]\,.
\ee
In this case, there are two conditions to be imposed on the energy-momentum tensor, namely the WEC and SEC.
The WEC is the least stringent one being this condition the one to be assumed for the FT to hold.
Hence, we are left with the following inequality
\begin{eqnarray}
R_{0}+f(R_{0}) \leq 0\,,
\label{ConstCurvSolutionIneq}
\end{eqnarray}
in order to get ${\cal M}_{\boldsymbol{\xi}} \leq 0$ where $G_{eff}>0$ was assumed. Hence, for constant scalar curvature spacetime this condition must be satisfied in order the FT holds provided the WEC is assumed on the energy-momentum tensor.

Nevertheless, as we have already noted, the FT does not hold at a cosmological scales in the present Universe \cite{ACD} since the deceleration parameter satisfies $q_{0}<0$. In this sense, one would like to know the necessary conditions for allowing \mbox{${\cal M}_{\boldsymbol{\xi}} > 0$} under the most stringent assumption on the energy-momentum tensor, namely the SEC, since under this condition the FT is accomplished in GR. For this purpose let us rearrange the field equations \eqref{fieldeq}
so the SEC appears explicitly. This may be done using \eqref{trace_R0} and the result is
\begin{eqnarray}
{\cal M}_{\boldsymbol{\xi}} = \frac{\frac{1}{2} \left( R_{0}\, f'(R_{0})-f(R_{0}) \right)-8\pi \left( T_{ab}\xi^a\xi^b + \frac{1}{2}T \right)}{1+f'(R_{0})}\,.
\end{eqnarray}
Then the assumption of both the SEC and $G_{eff}>0$ implies
\begin{eqnarray}
{\cal M}_{\boldsymbol{\xi}} \leq \frac{R_{0}\, f'(R_{0})-f(R_{0}) }{2\left(1+f'(R_{0})\right)}\,.
\end{eqnarray}
Then, a necessary condition to get ${\cal M}_{\boldsymbol{\xi}} > 0$ is
\begin{eqnarray}
R_{0}\, f'(R_{0})\ - f(R_{0}) > 0\,.
\end{eqnarray}
This condition together with the assumption of vacuum\footnote{This assumption tries to
encapsulates the Universe late-time epoch evolution when matter density is sufficiently diluted and negligible with respect to de Sitter final stages.}
was used in \cite{ACDD} to impose several restriction on the parameters of different $f(R)$ models in order to guarantee
cosmological acceleration.


\section{Friedmann-Robertson-Walker in $f(R)$ theories}
\label{Sec:RW-f(R)}

In this section we shall generalize the conditions guaranteeing the FT for $f(R)$ theories in FRW spacetimes. These spacetimes possess
non-constant scalar curvature and from the field equations for $f(R)$ theories, 
we can obtain the generalized Friedmann equation (GFE) for a perfect fluid yielding
\be
\partial^2_{t} f' 
\,=\,2H^2(1+f')-\frac{1}{3}[R+f+8 \pi G(\rho+3 p)]
\label{GFE}
\ee
as well as the well-known temporal component of the equations of motion
\be
\dot{\rho}
= -3 H (\rho + p).
\ee
%


\subsection{Timelike Geodesics}


For spacetimes with non-constant curvature solutions,
the curvature contribution to the Raychaudhuri equation translates into
\begin{eqnarray}
{\cal M}_{\boldsymbol{\xi}}=\frac{\frac{1}{2}\left( R+f \right)-h^{ab}\nabla_{a}\nabla_{b}f'-8\pi T_{ab}\xi^a\xi^b}{1+f'}\,,
\label{Raycon}
\end{eqnarray}
where $ h_{ab}\equiv g_{ab}+\xi_a\xi_b$   is the projector onto the subspace orthogonal to $\boldsymbol{\xi}$.

Provided the WEC is assumed we obtain
\begin{eqnarray}
{\cal M}_{\boldsymbol{\xi}}\leq \frac{\frac{1}{2}\left( R+f \right)-h^{ab}\nabla_{a}\nabla_{b}f'}{1+f'}\,.
\end{eqnarray}
For FRW models and according to results in Section
\ref{Sec:RW} the last expression can be simplified yielding
\begin{eqnarray}
{\cal M}_{\boldsymbol{\xi}}\leq \frac{\frac{1}{2}\left( R+f \right)+(1-\gamma^2)\partial^2_{t}f'+\left( \frac{\gamma_0^2 v_0^2 a_0^4}{a^2}+3 \right) H \partial_{t} f'}{1+f'}\,.
\end{eqnarray}
Since $\gamma\geq 1$, sufficient conditions for the FT theorem to hold
are
\begin{eqnarray}
R + f \leq 0 \ \ \ \ \ \partial^2_{t}{f}' \geq 0 \ \ \ \ \ H \partial_{t} f' \leq 0\,.
\end{eqnarray}
This result reduces to (\ref{ConstCurvSolutionIneq}) provided that constant curvature solutions are considered.

On the other hand, if we are interested in assuming the SEC the equation \eqref{Raycon} may be recast into
\begin{eqnarray}
{\cal M}_{\boldsymbol{\xi}}\,&=&\,\frac{\frac{1}{2}\left( f'\,R-f \right)-h^{ab}\nabla_{a}\nabla_{b}f'+\frac{3}{2}\Box f'}{1+f'} \nonumber \\
&-&\frac{8\pi\left( T_{ab}\xi^a\xi^b+\frac{1}{2}T\right)}{1+f'}\,.
\label{SEC_0_fR}
\end{eqnarray}
Therefore after assuming the SEC, (\ref{SEC_0_fR}) becomes
\begin{eqnarray}
{\cal M}_{\boldsymbol{\xi}}\leq\frac{\frac{1}{2}\left( f'\,R-f \right)-h^{ab}\nabla_{a}\nabla_{b}f'+\frac{3}{2}\Box f'}{1+f'}\,.
\end{eqnarray}
In a FRW model, this inequality can be expressed as
\begin{eqnarray}
{\cal M}_{\boldsymbol{\xi}}&\leq& \frac{\frac{1}{2}\left( f'\,R-f \right) - \left( \frac{1}{2}+\gamma^2 \right) \partial^2_{t} f'}{1+f'}\,\nonumber\\
&+&\frac{\left( \frac{\gamma_0^2 v_0^2 a_0^4}{a^2}-\frac{3}{2}\right) H \partial_{t}f'}{1+f'}\,.
\end{eqnarray}
Thus, sufficient conditions for ${\cal M}_{\boldsymbol{\xi}}<0$ assuming the SEC would be
\begin{eqnarray}
f'\,R-f \leq 0\ \ \ \ \ \ \partial^2_{t} f' \geq 0\,
\label{Suf_Conds_1_timelike}
\end{eqnarray}
and
\begin{eqnarray}
&\text{if}\ \ \  \frac{\gamma_0^2 v_0^2 a_0^4}{a^2} \geq \frac{3}{2} \ \ \ \text{then}\ \ \ H\partial_{t} f' \leq 0 \nonumber \\
&\ \text{if}\ \ \  \frac{\gamma_0^2 v_0^2 a_0^4}{a^2} \leq \frac{3}{2} \ \ \ \text{then}\ \ \ H\partial_{t} f' \geq 0\,.
\label{Suf_Conds_2_timelike}
\end{eqnarray}
However, it is not possible to derive sufficient conditions for the FT theorem to hold since the last condition cannot be satisfied simultaneously for all timelike geodesics.
%


\subsection{Null Geodesics}


The ${\cal M}_{\boldsymbol{k}}$ expression for null geodesics reads
\be
 {\cal M}_{\boldsymbol{k}} = -\frac{k^a k^b \nabla_{a}\nabla_{b}f'+8\pi T_{ab}k^a k^b}{1+f'}\,.
\ee
Assuming the NEC, we obtain
\be
 {\cal M}_{\boldsymbol{k}} \le -\frac{k^a k^b \nabla_{a}\nabla_{b}f'}{1+f'}\,,
\ee
and considering FRW spacetime it follows
\begin{eqnarray}
{\cal M}_{\boldsymbol{k}} \le -\frac{\frac{k_{0}^{2}a_{0}^{2}}{a^2}\left( \partial_{t}^2 f' - H \partial_{t} f' \right)}{1+f'}\,.
\end{eqnarray}
Thus,
\begin{eqnarray}
\partial^2_{t}{f}' \geq 0 ,\ \ \ \ \ H \partial_{t} f' \leq 0\,.
\label{Suf_Cond_null}
\end{eqnarray}
turns out to be a sufficient condition to ensure the FT theorem for null geodesics.

%
%

\section{The case of  $R+f(R) \propto R^{n}$  theories}

In order to a better understanding of the possibilities to applying the above CC
leading to the FT it is interesting to study relevant spacetimes.
In this sense, the reconstruction methods in $f(R)$ gravity theories
try to invert the field equations to deduce what class of
$f(R)$ theories give rise to a particular cosmological expansion history \cite{18}.
For phases of cosmic evolution  when the energy density is
dominated by a perfect fluid, e.g., dust or radiation, the scale factor can be represented by
exact power-law solutions of the form
\be
a_{}(t)\,=\,a_0\left(\frac{t}{t_0}\right)^{m_{}}.
\label{scale factors power laws}
\ee
In \cite{19} the authors found that such expansion histories exist only for modifications where the total gravitational Lagrangian is a simple power  of R, i.e.,
 $R+f(R) \propto R^{n}$ whenever matter is described
by a single component perfect fluid with an equation of state given by $p= \omega \rho$ with constant $\omega$.
In this scenario,  the exponent $m $ satisfies
\be m=\frac{2n}{3(1+ \omega)}. \ee
Moreover the integration of the field equations proved that an dynamical-system orbit representing a cosmic history passes close to the
matter dominated point $G$ and eventually tends toward
the late-time attractor $C$ \cite{17} where provided that $1.36 <n<1.5$, $G$ and $C$, respectively, represent a decelerated
matter-dominated and late-time power-law acceleration solutions.
As a matter of fact, let remind that $R^n$ gravities are severely constrained by
solar system tests \cite{50} as well as by the evolution baryon acoustic oscillations and  scalar and tensor perturbations \cite{29}.
In the following, let us assume $n$ to be a  positive integer number, $n=1,2,3...$ so that $n=1$
is just GR (with $\lambda_1= 1$).  In these models the scalar
curvature will be negative $R<0$ provided
\be \frac{n}{1+w}  <
\frac{3}{4} \ee
and then the CC for null geodesic congruences reads
\be
m \ge 0 . \ee
Now assuming $w \in (-1,1]$ we get the result that for
any natural number $n > 0$, so that the CC is fulfilled and the
FT applies.

For timelike geodesics  the CC is given by
\be
\gamma_0^2v_0^2a_0^2\left(\frac{t}{t_0}\right)^{2m}  \ge \frac{3}{2}(m -1). \ee
Since $\gamma_0^2v_0^2a_0^2 \in [0, \infty)$, it is clear
that the CC and consequently the FT apply for $n\ge  3$.

For the case
$n=2$ it is trivial to check that for $w \ge 1/3$, i.e. for $T \le 0$, again
both the CC and FT are satisfied. However, for $w < 1/3$ it is possible to show
that  for any given initial condition there is a $t^*$ so that for any $t>
t^*$ the CC is violated and therefore there is no FT.
For example for the case of dust (i.e., $w=0$) the CC reads
\be
\gamma_0^2v_0^2a_0^2(t/t_0)^{8/5} \ge \frac{1}{2}
\ee
 and it is clear that the FT will be violated at large times.

 For the $n=1$ case the CC renders the condition
\be
\gamma_0^2v_0^2a_0^2\left(\frac{t}{t_0}\right)^{\frac{4}{3(1+w)}}  \ge
-\frac{1+3w}{2(1+w)}.
\ee
The above condition is manifestly fulfilled
whenever $1+3\omega  \ge 0$ but this is just the SEC. In other words,
the last expression  allows to reobtain
the results presented at the beginning of this investigation for GR. If on the
contrary  $1+3\omega  < 0$ then again for any initial condition there exists
a given time $t^*$ such that the CC
is not  valid for times $t>t^*$.

Thus, for $R^n$ gravities we have FT in all cases except for timelike geodesics with
 $n=2$ provided that $\omega<1/3$ and $n=1$ provided that $1+3\omega <0$.


\section{Two $f(R)$ paradigmatic models}

In order to illustrate our results we have considered two general classes for the $f(R)$ gravitational Lagrangian.
The first one is a polynomial of the form
\be
R+f(R)\,=\,H_0^2\left[a_p \frac{R}{H_0^2}+ \ b_p \left(\frac{R}{H_0^2}\right)^{n_p}\right]
\label{Model_1}
\ee
where 
$\{a_p,\, b_p,\, n_p\}$ are constants. These Lagrangians may be regarded as
natural extension to GR. The literature devoted to such models is extensive depending on the values of the exponent $n$ under consideration \cite{fR}.
The considered values $a_p=-0.5$, $b_p=0.92$ and $n_p=1.2$ provided a good fit to the Hubble parameter today ($h_0=1.1011$) albeit the obtained value for
$q_0=-0.0356$ does not provide acceleration today. This class of models must be then considered as an illustrative example.

\vspace{0.1cm}
The second model under consideration is the well-known Hu-Sawicki model \cite{Sawicki}. This model has been claimed as viable in the cosmological realm and
several tests are available in literature \cite{Sawicki_citations}. The Lagrangian of the aforementioned model can be written as
\be
f(R)\,=\, -m^2\frac{a_1^{HS}\left(\frac{R}{m^2}\right)^{n^{HS}}}{1+a_2^{HS}\left( \frac{R}{m^2}\right)^{n^{HS}}}\,,
\label{Model-f(R)-HS}
\ee
where $a_{1,2}^{HS}$ and $n_{HS}$ represent dimensionless parameters, $m^2$ may be understood as a coupling constant
possessing dimensions of Hubble parameter squared. The considered values were $a_1^{HS}  = 1$, $a_2^{HS} = 1/19$, $n^{HS} = 2$
and $m^2 = \frac{6}{19}H_0^2(1-\Omega_m^0)$ in order to guarantee the $\Lambda$CDM recovery at high curvatures as well as to provide
a good fit to both Hubble and deceleration parameters today ($h_0 = 0.9983$ and $q_0 = -0.5274$).

\vspace{0.1cm}

\begin{figure*}[htbp] 
	\centering
		\includegraphics[width=0.4250275\textwidth]{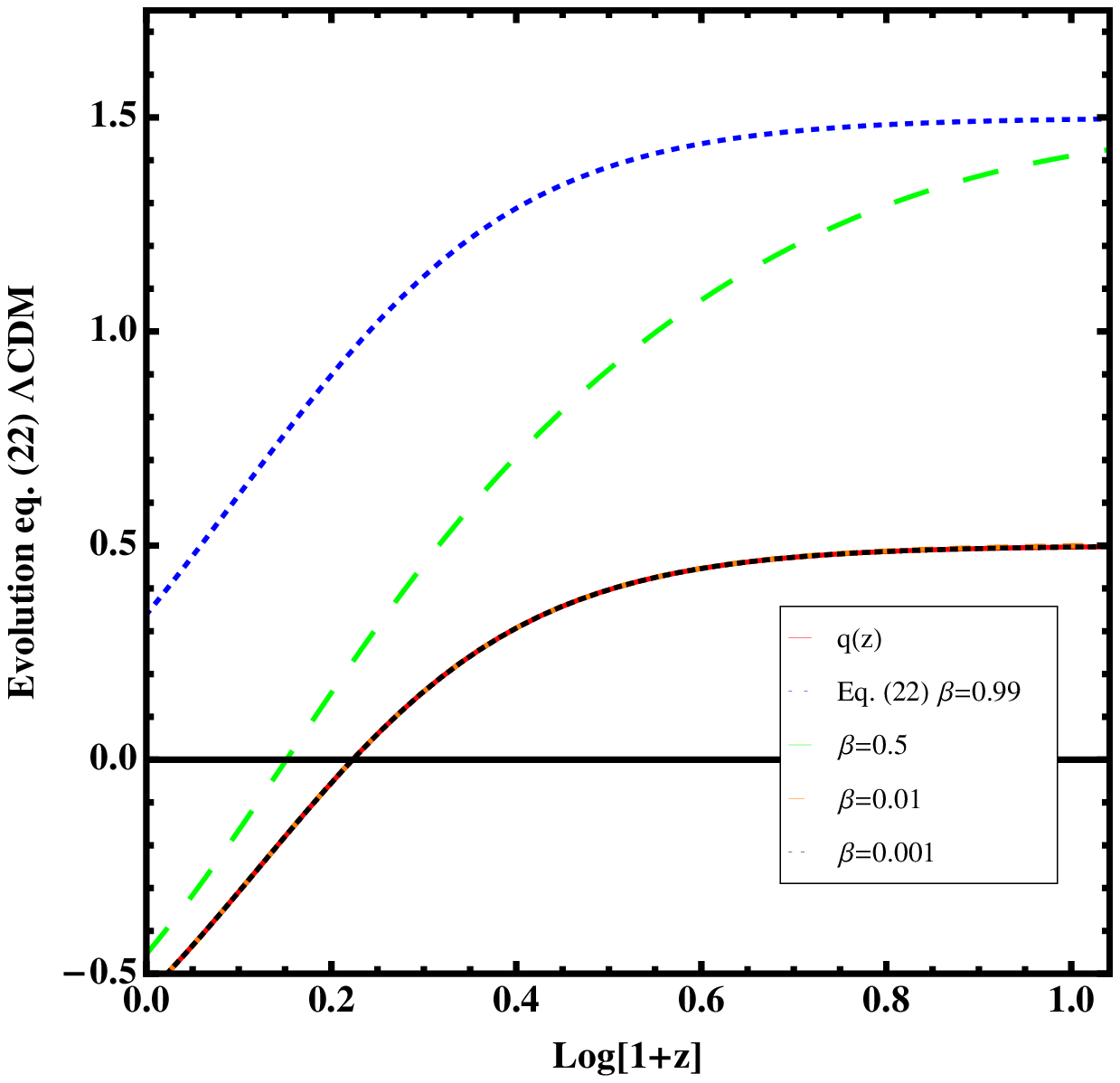}
		\includegraphics[width=0.4250275\textwidth]{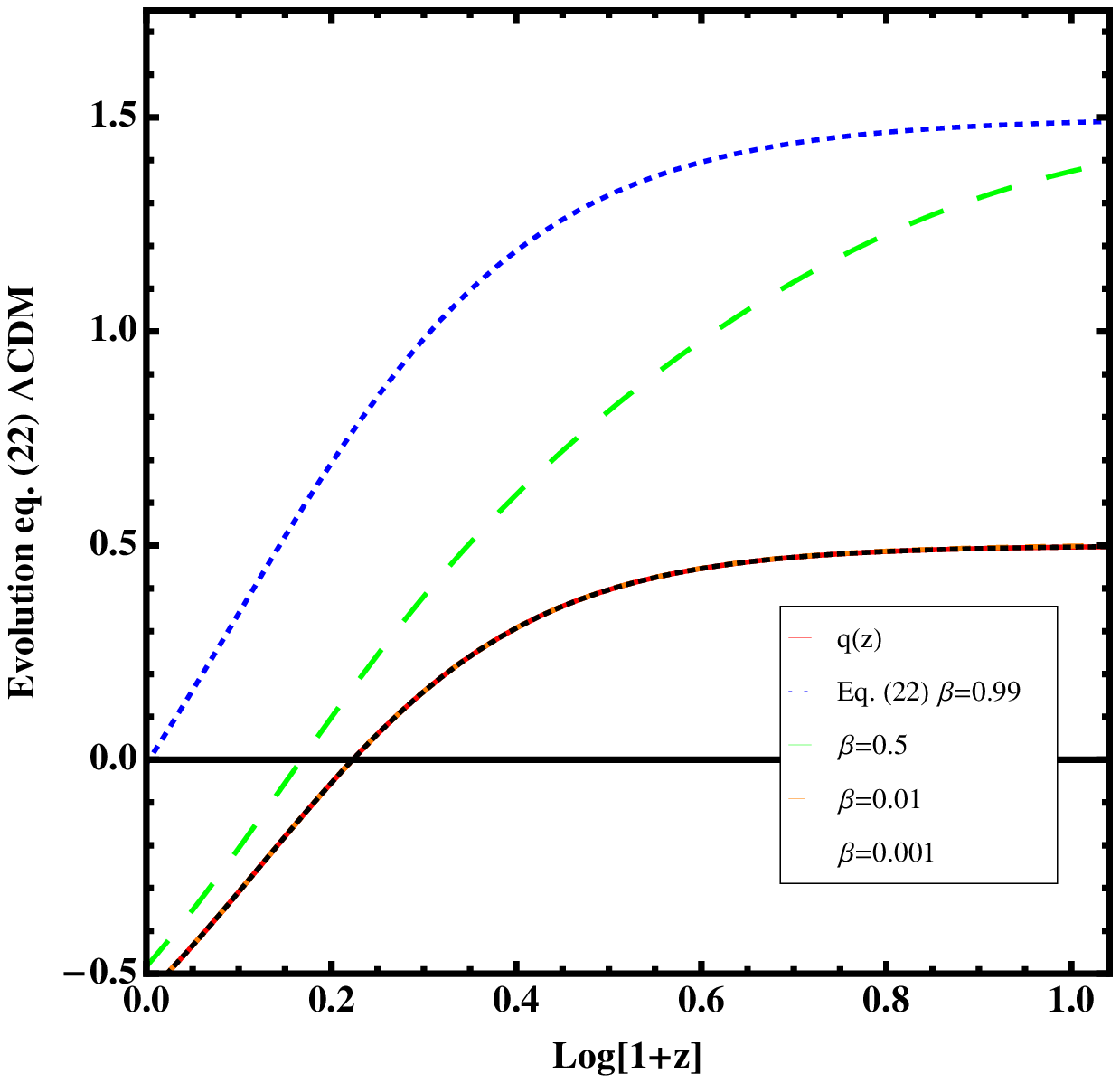}\\
		 \includegraphics[width=0.4250275\textwidth]{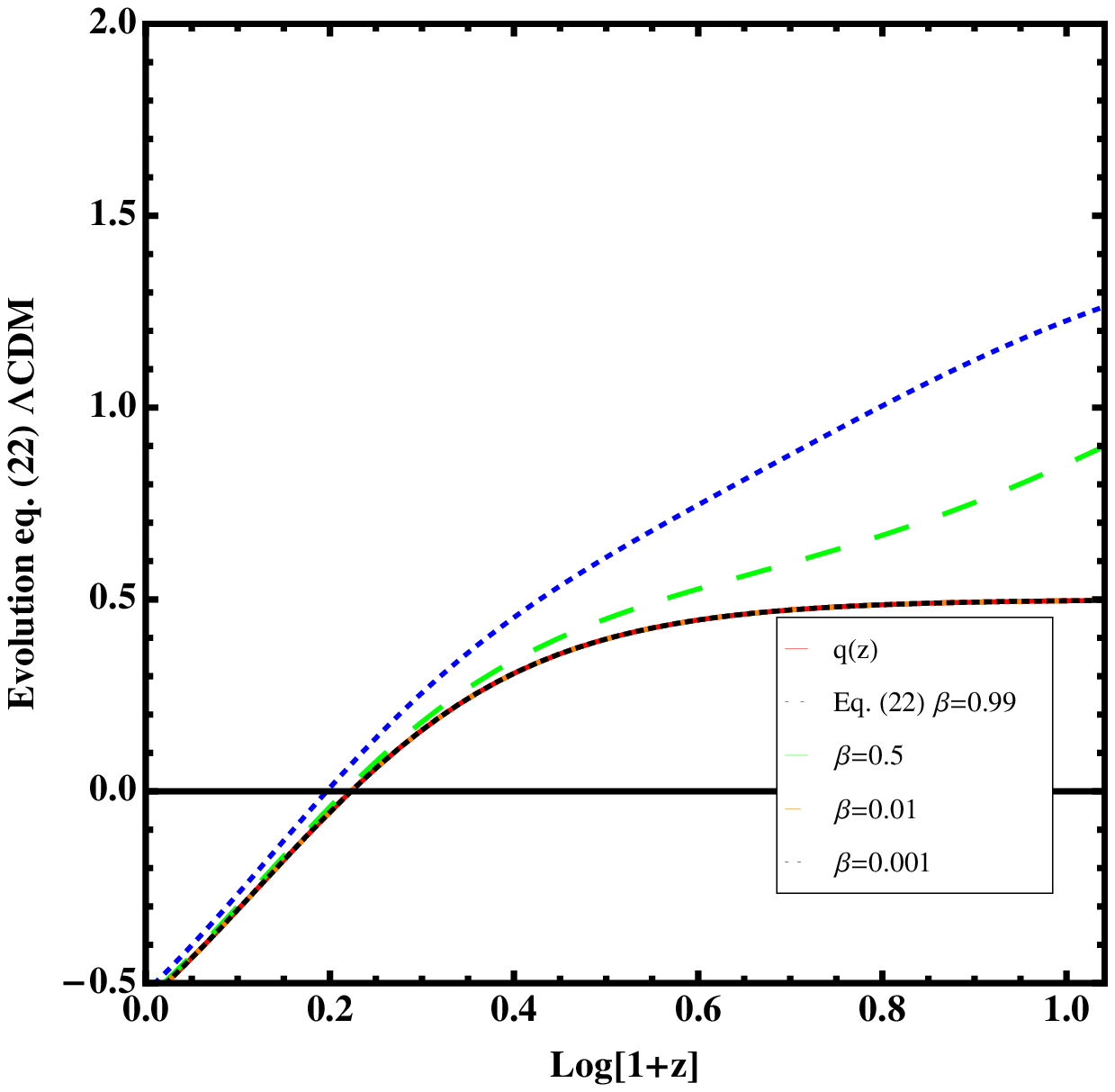}
		\includegraphics[width=0.4250275\textwidth]{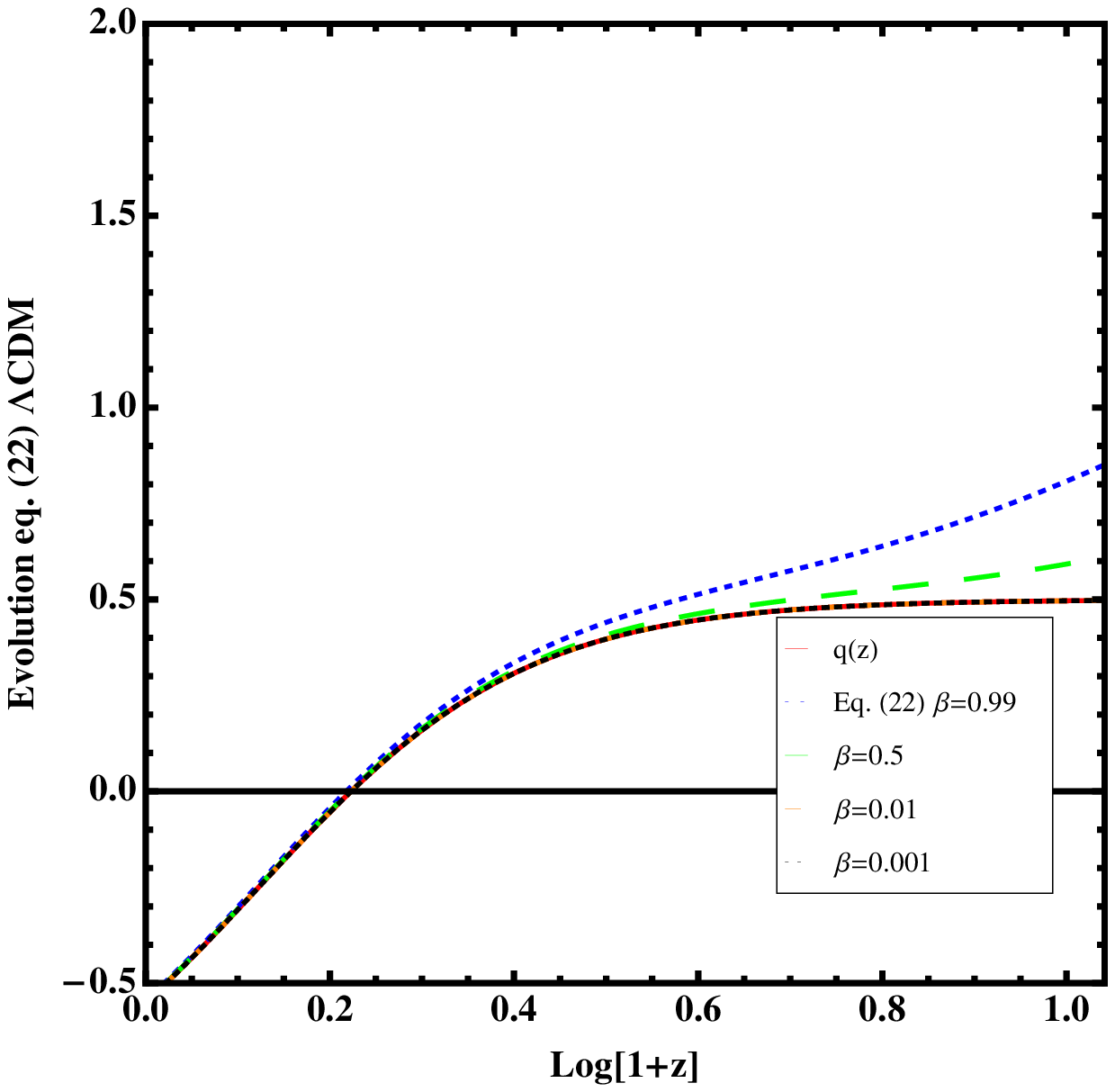}
 \caption{\footnotesize{
Evolution in redshift of equation (\ref{CCtime_2}) l.h.s for $\Lambda$CDM model with $\Omega_m=0.3$, $\Omega_{\Lambda}=0.7$:
In every panel both the deceleration parameter $q(z)$ and l.h.s. of equation  (\ref{CCtime_2}) are depicted.
The redshift interval was $[0,\,10]$.
Different initial velocities
$\beta_0=\{0.99,\, 0.5,\, 0.01,\, 0.001\}$ on the geodesic at given redshifts are plotted in every panel. The initial redshift  (time) at which the motion on the geodesics is
considered to start is different for every panel. In particular
$z_0 = 0.01$ (upper left),
$z_0 = 0.1$ (upper right),
$z_0 = 1$ (lower left)
and
$z_0 = 2$ (lower right).
For small initial velocities $\beta_0$ one can see that the evolution for the
 l.h.s. of equation (\ref{CCtime_2}) almost coincides with the $q(z)$ evolution an therefore the lines are essentially indistinguishable. For intermediate and relativistic initial velocities ($\beta_0  \sim 1$)
 the second term in l.h.s of equation (\ref{CCtime_2}) increases the obtained value. Thus, even if $q(z)$ was negative at some stages of the cosmological evolution, the  l.h.s. of equation  (\ref{CCtime_2}) becomes positive and therefore the Focusing Theorem is eventually satisfied.
 }}
 \label{Figure_Eq_39_LCDM}
\end{figure*}
\begin{figure*}[htbp] 
	\centering
		\includegraphics[width=0.4250275\textwidth]{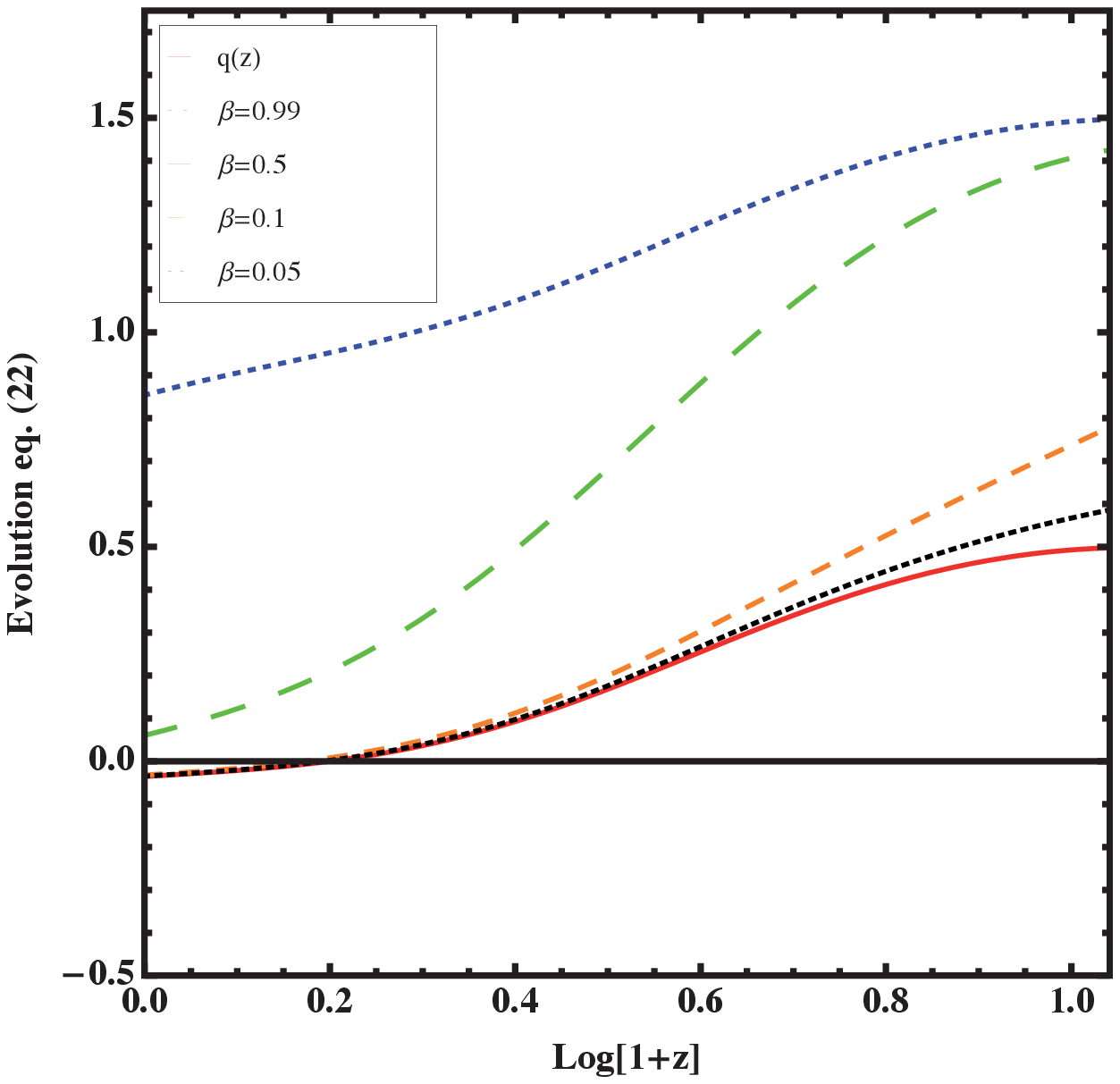}
		\includegraphics[width=0.4250275\textwidth]{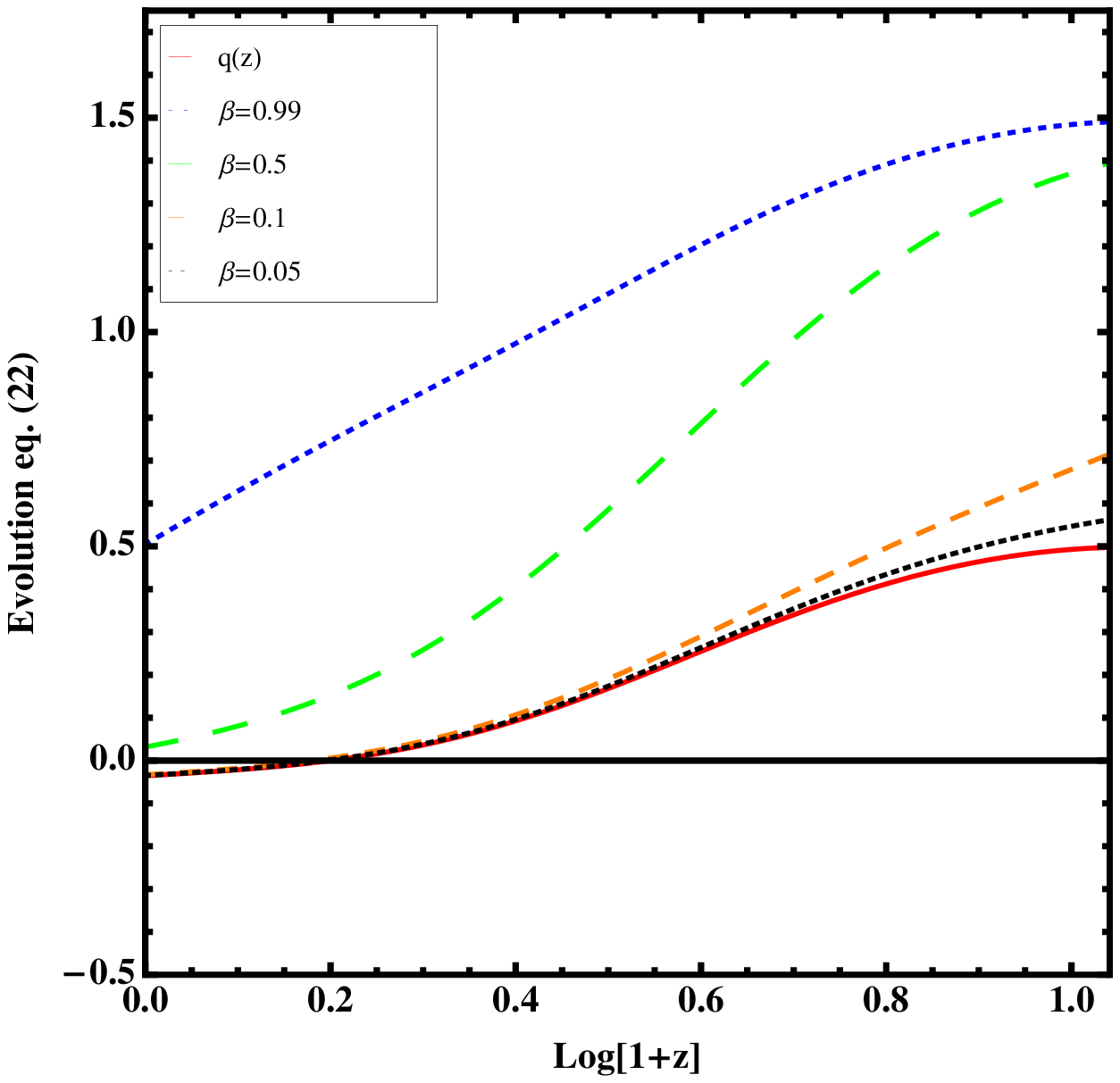}\\
		 \includegraphics[width=0.4250275\textwidth]{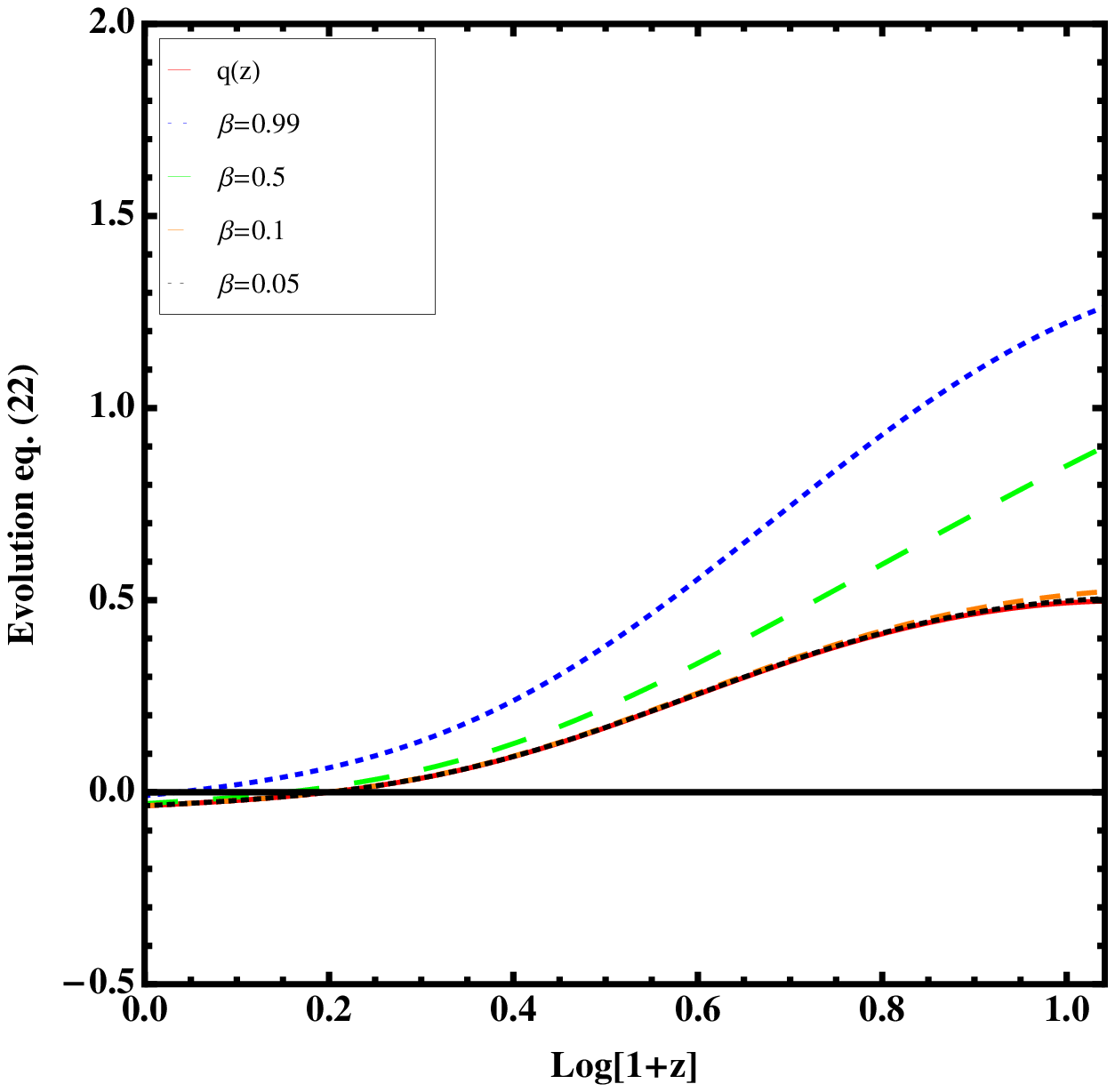}
		\includegraphics[width=0.4250275\textwidth]{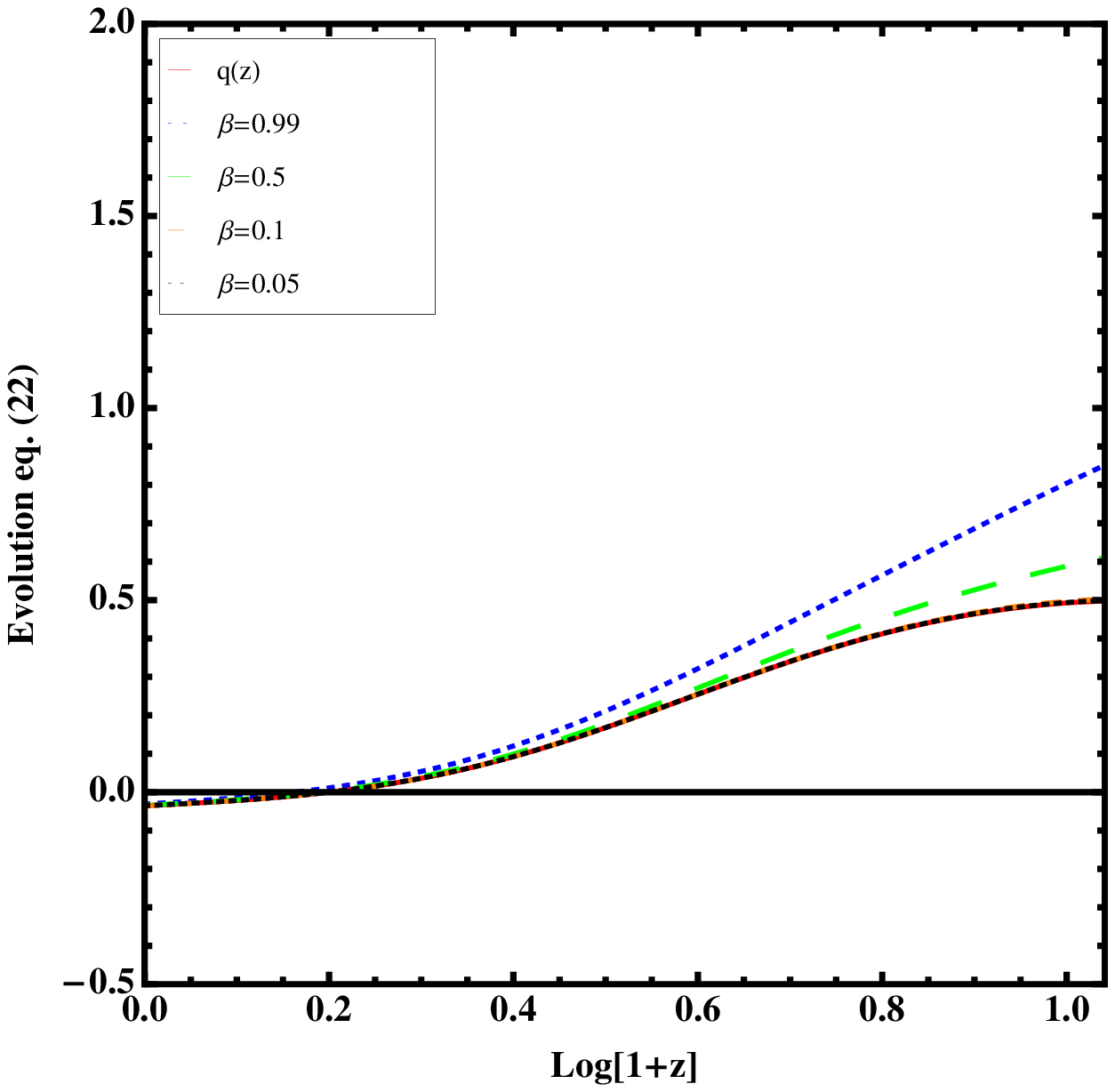}
 \caption{\footnotesize{
Evolution in redshift of equation (\ref{CCtime_2}) l.h.s for $f(R)$ model given in 
(\ref{Model_1}) with values $a_p=-0.5$, $b_p=0.92$ and $n_p= 1.2$. Initial conditions for this model were given at $z=10$ and coincided
 with $\Lambda$CDM at this redshift.
In every panel the deceleration parameter $q(z)$ and l.h.s. of equation  (\ref{CCtime_2}) are depicted.
The redshift interval was $[0,\,10]$.
Different initial velocities
$\beta_0=\{0.99,\, 0.5,\, 0.01,\, 0.001\}$ on the geodesic at given redshifts are depicted in every panel. The initial redshift  (time) at which the motion on the geodesics is
considered to start is different for every panel. In particular
$z_0 = 0.01$ (upper left),
$z_0 = 0.1$ (upper right),
$z_0 = 1$ (lower left)
and
$z_0 = 2$ (lower right). The analysis is completely analogous to the results presented in Fig.  \ref{Figure_Eq_39_LCDM}
whilst specific evolutions and redshift for sign reversals may differ.
 %
%
 }}
 \label{Figure_Eq_39_Polynomial}
\end{figure*}
%
%
%
%
\begin{figure*}[htbp] 
	\centering
		\includegraphics[width=0.4250275\textwidth]{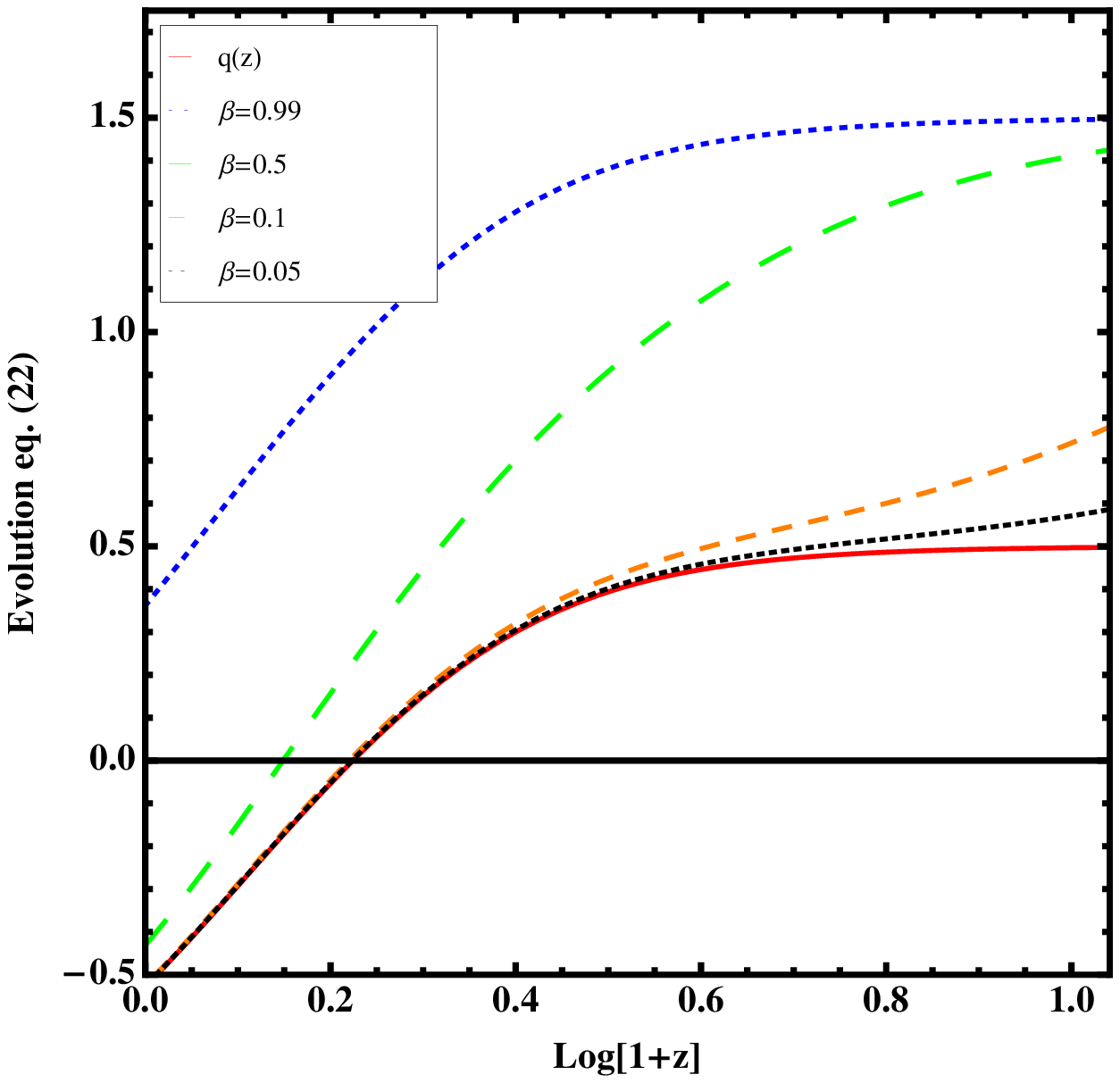}
		\includegraphics[width=0.4250275\textwidth]{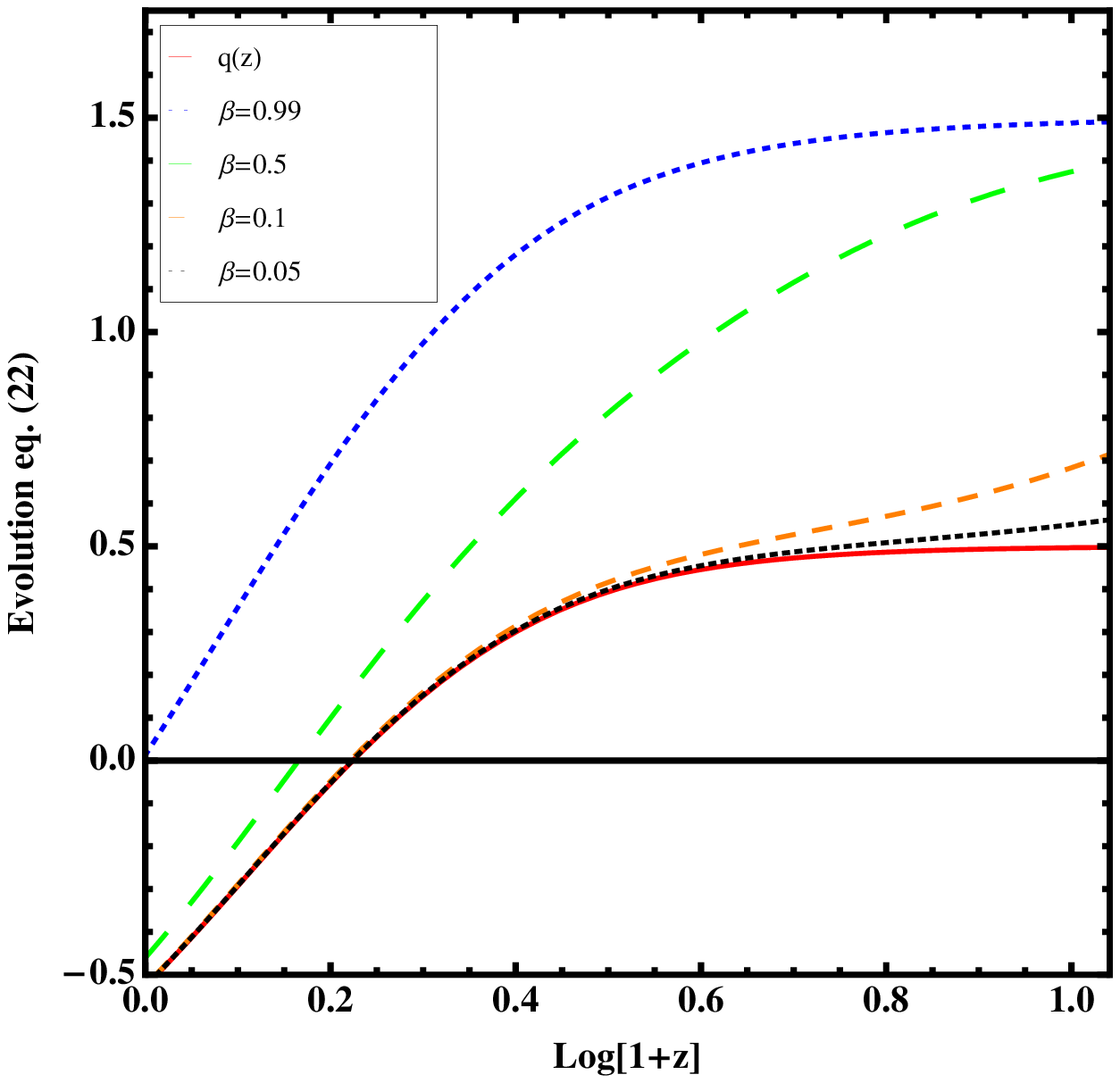}\\
		 \includegraphics[width=0.4250275\textwidth]{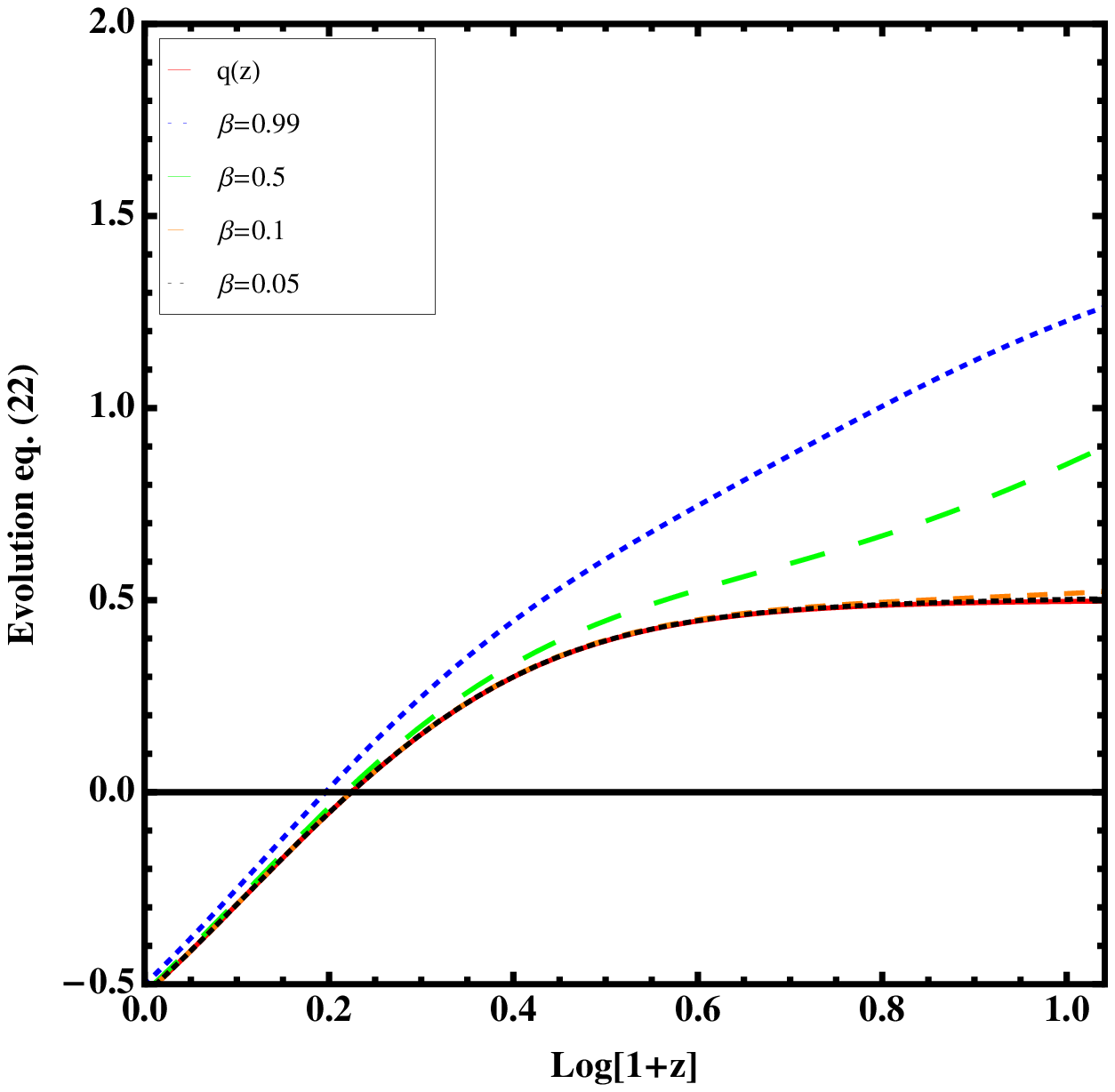}
		\includegraphics[width=0.4250275\textwidth]{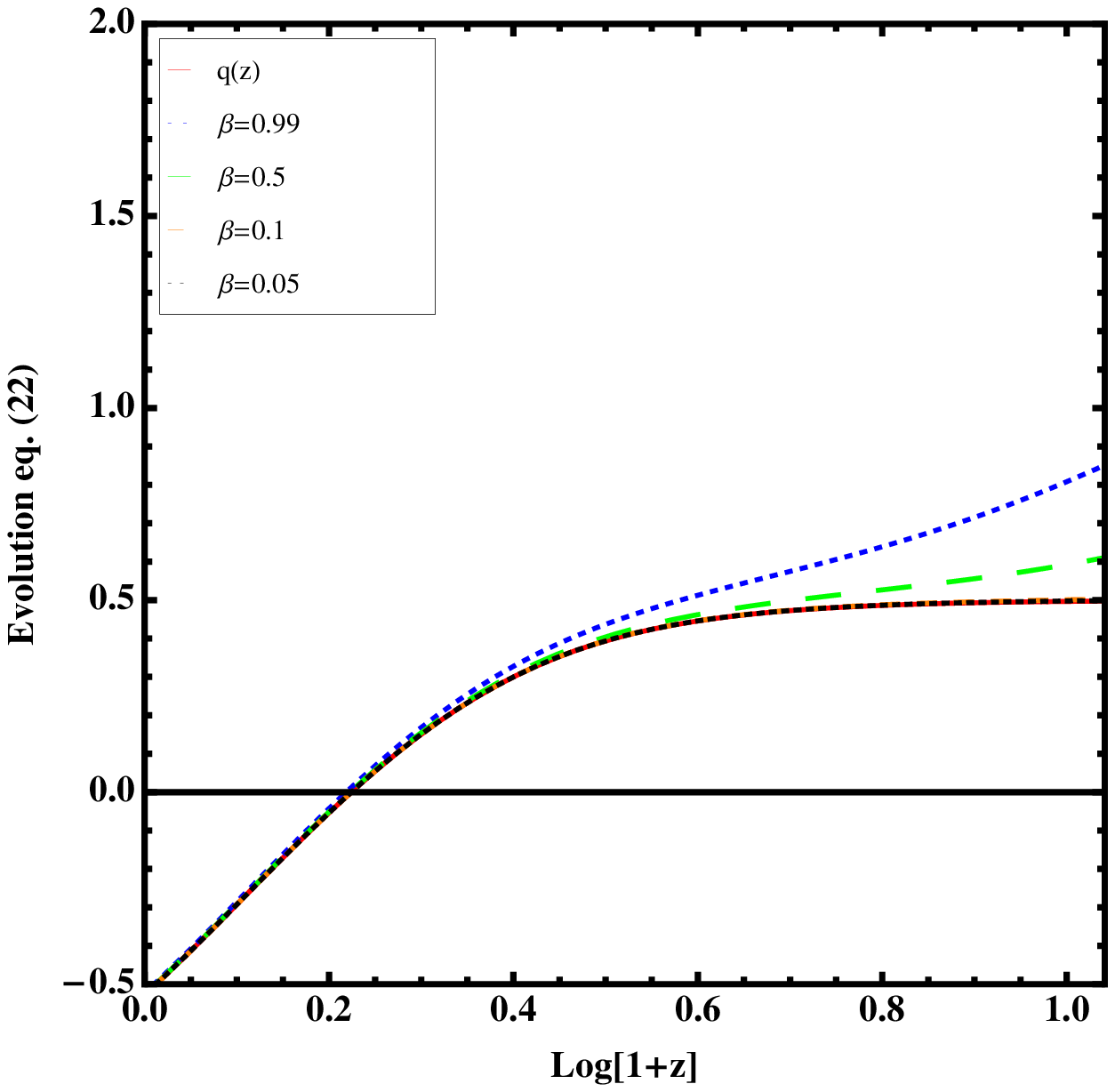}
 \caption{\footnotesize{
Evolution in redshift of equation (\ref{CCtime_2}) l.h.s for $f(R)$ model given in (\ref{Model-f(R)-HS}) with values
$a_1^{HS}  = 1$, $a_2^{HS} = 1/19$, $n^{HS} = 2$ and $m^2 = \frac{6}{19}H_0^2(1-\Omega_m^0)$.
%
 Initial conditions for this model were given at $z=10$ and coincided with $\Lambda$CDM at this redshift.
In every panel the deceleration parameter $q(z)$ and l.h.s. of equation  (\ref{CCtime_2}) are depicted.
The redshift interval was $[0,\,10]$.
Different initial velocities
$\beta_0=\{0.99,\, 0.5,\, 0.01,\, 0.001\}$ on the geodesic at given redshifts are depicted in every panel. The initial redshift  (time) at which the motion on the geodesics is
considered to start is different for every panel. In particular
$z_0 = 0.01$ (upper left),
$z_0 = 0.1$ (upper right),
$z_0 = 1$ (lower left)
and
$z_0 = 2$ (lower right). The analysis is completely analogous to the results presented in Fig.  \ref{Figure_Eq_39_LCDM}
whilst specific evolutions and redshift for sign reversals may differ.
 %
%
 }}
 \label{Figure_Eq_39_HuSawicki}
\end{figure*}

 In Figs. \ref{Figure_Eq_39_LCDM},  \ref{Figure_Eq_39_Polynomial} and \ref{Figure_Eq_39_HuSawicki}
 we have depicted the evolution of the left-hand side (l.h.s.) of equation (\ref{CCtime_2}) for both $\Lambda$CDM and $f(R)$ models of the form (\ref{Model_1}) and
 (\ref{Model-f(R)-HS}). One can see how depending on the initial redshift and velocity of the geodesics the  l.h.s. of equation  (\ref{CCtime_2}) becomes positive at some stages of the cosmological evolution and therefore the Focusing Theorem is eventually satisfied.


\section{Conclusions}
\label{sec:conclusions}

In this paper we have derived the convergence condition for both timelike and null geodesics in spatially flat Robertson-Walker spacetimes.
For null congruences the accomplishment of this condition
is determined solely by the deceleration parameter whereas for timelike congruences
it depends upon both the deceleration factor, the redshift and the velocity at which the first integral given by expression (\ref{CCtime_2}) is evaluated.
Thus the combination of such terms may lead to either violation or accomplishment of the Focusing Theorem for given geodesics in different cosmological epochs depending
on the underlying gravity theory. At this stage let remind that for the accepted values of the $\Lambda$CDM Concordance model in a Robertson-Walker spacetime, the convergence condition is violated at late times and thus the Focusing Theorem does not hold.

We then have applied these results to two paradigmatic classes of fourth-order gravity theories, namely a polynomial function of the Ricci scalar and the so-called Hu-Sawicki model.
For these classes of models we established a comparison with the well-known
$\Lambda$CDM results. The qualitative behavior for timelike geodesics indicates that the higher initial redshift or the smaller the initial velocity, the departure of expression (\ref{CCtime_2})
with respect to the value of the deceleration parameter is smaller.
Since the accomplishment of the convergence condition for null geodesics is determined solely by the deceleration parameter, we conclude that
depending on the model under consideration, timelike and null geodesics
would evolve differently depending on the initial velocity and redshift.

Also in the frame of the fourth-order gravity theories in the metric formalism we have obtained sufficient conditions that such theories must accomplish in order to guarantee the convergence condition for
timelike and null geodesics. For the timelike geodesics we proved that it is not possible to derive full sufficient conditions for the Focusing Theorem to hold since one
of the two required conditions -  (\ref{Suf_Conds_2_timelike}) - cannot be satisfied for all the timelike geodesics. On the contrary, for the null geodesics it was possible to derive two sufficient conditions as given in (\ref{Suf_Cond_null}).
This analysis can be easily applied to competitive fourth-order gravity models claimed as valid to study the geodesics evolution behavior.

We have analyzed the convergence condition for paradigmatic power-law solutions whose importance in several cosmological stages is manifest. The accomplishment of the convergence condition was proved to depend either solely on the exponent of the power law (null geodesics) or both the exponent and the initial parameters of the geodesics (timelike geodesics).

Finally, we have extended the analysis to anisotropic geometries. In particular, we have shown that Kasner spacetimes provide simple examples of power-law evolutions where the convergence condition is not fulfilled both for null and timelike geodesics.


\subsection{Acknowledgments}
This work has been supported by MICINN (Spain) projects numbers
FIS2011-23000, FPA2011-27853-C02-01, FPA2011-27853-C02-01, UCM-BSCH GR58/08 910309 and Consolider-Ingenio MULTIDARK CSD2009-00064.
F.D.A. acknowledges financial support from the UAM+CSIC Campus of International Excellence (Spain).
A.d.l.C.D. thanks the financial support from ACGC University of Cape Town and IEEC-ICE Barcelona in 2013 as well as 
the hospitality of Kavli Institute for Theoretical Physics China (KITPC) during the preparation of this manuscript.
%


\end{document}